 \documentclass[final,3p,times,authoryear]{elsarticle}

\usepackage{amssymb}

\usepackage{color}
\usepackage{colortbl}
\usepackage{multirow}
\usepackage{todonotes}
\usepackage{relsize,etoolbox}%
\usepackage{float}
\usepackage{url}
\usepackage{multicol}
\usepackage{enumitem}
\usepackage{lscape}
\usepackage{booktabs} 
\usepackage{comment}
\usepackage{lipsum}
\usepackage{array}
\usepackage[symbol]{footmisc}
\usepackage{xcolor}

\usepackage{natbib}

\journal{International Journal of Human Computer Studies}

\begin{document}
 
\pagenumbering{arabic}

\begin{frontmatter}



\title{Talking datasets -- understanding data sensemaking behaviours}



\author[label1]{Laura Koesten*}

\author[label2]{Kathleen Gregory*}


\author[label3]{Paul Groth} 

\author[label1]{Elena Simperl}

\address[label1]{King's College London}
\address[label2]{Data Archiving and Networked Services, Royal Netherlands Academy of Arts \& Sciences}
\address[label3]{University of Amsterdam, NL}

\begin{abstract}
The sharing and reuse of data are seen as critical to solving the most complex problems of today. Despite this potential, relatively little attention has been paid to a key step in data reuse: the behaviours involved in data-centric sensemaking. We aim to address this gap by presenting a mixed-methods study combining in-depth interviews, a think-aloud task and a screen recording analysis with $31$ researchers from different disciplines as they summarised and interacted with both familiar and unfamiliar data. We use our findings to identify and detail common patterns of data-centric sensemaking across three clusters of activities that we present as a framework: \textit{inspecting} data, \textit{engaging} with content, and \textit{placing} data within broader contexts. Additionally, we propose design recommendations for tools and documentation practices, which can be used to facilitate sensemaking and subsequent data reuse.

\end{abstract}




\end{frontmatter}


\section{Introduction}
\label{introduction}

\renewcommand{\thefootnote}{\fnsymbol{footnote}}
\textcolor{white}{\footnote{Both authors contributed equally to this research.}}

Climate change; poverty; global hunger - all have been dubbed wicked problems \citep{peters2017so} that have the best chance of being tackled by bringing together and using cross-disciplinary data in new ways~\citep{walshe2020introduction}.{\setcounter{footnote}{0} \renewcommand*{\thefootnote}{\arabic{footnote}}\footnote{See CODATA Decadal Program \url{ http://www.codata.org/strategic-initiatives/decadal-programme}.}\setcounter{footnote}{1} \renewcommand*{\thefootnote}{\arabic{footnote}}\footnote{See United Nations on Big Data and Development \url{https://www.un.org/en/sections/issues-depth/big-data-sustainable-development}.}} Although data reuse is increasingly encouraged \citep{EUpublications}, it involves a host of challenges, such as providing rich, standardised metadata adequate for interoperability and reuse  \citep{wilkinson2016fair}. Fundamentally, reusing data also requires that data consumers make sense of data that others have created. \newline 

Even within their own disciplinary domains, understanding and making sense of data is a difficult and time--intensive process for researchers and data professionals \citep{kern2015there,DBLP:conf/chi/MullerLWPTLDE19} which is heightened by the demands of navigating an increasing amount of digital information \citep{doi:10.1080/01972240490507974}. Also contributing to this difficulty is the fact that data do not speak for themselves, but require supporting structures -- both social and technical -- to convey the meaning necessary for reuse \citep{borgman2015big}. 
The effort and costs involved in sensemaking can potentially be reduced through the development of automated tools and systems \citep{DBLP:conf/chi/RussellSPC93}. Designing such tools is contingent upon first understanding and describing the behaviours involved in data-centric sensemaking \cite{DBLP:books/daglib/0029131}. \newline

Here, we identify and detail {\em patterns of activities} involved in data exploration and sensemaking. In the context of this work, data can be thought of as collections of related observations organised and formatted for a particular purpose, reflecting the variety of concepts different actors have of data, see \citep{borgman2015big}. In order to identify these patterns, we 
draw on verbal summarizations as a method of uncovering sensemaking processes and build on the following ideas: (1) the act of summarizing is a form of sensemaking, (2) verbal summarization represents unique cognitive processes and (3) it is possible to identify common patterns in sensemaking activities when people describe data that they are familiar with and data that are unknown to them. We used these ideas to develop the following research questions:

\begin{itemize}
\item RQ1: What are common patterns of activities, both for known and for unknown data, in the initial phases of data-centric sensemaking? 
\item RQ2: How do patterns of data-centric sensemaking afford potential data reuse?
\end{itemize}

To explore these questions, we combined in-depth interviews with researchers, in which they performed a think-aloud task, and a screen recording analysis. During the interviews, researchers interacted with and verbally summarised an example of their own research data and a dataset that was unknown to them. We present the results from this study and use our findings to identify activity patterns and data attributes which are important across three clusters of sensemaking activities: \textit{inspecting} data, \textit{engaging} with data content more deeply and \textit{placing} data within broader contexts. Finally, we detail design recommendations for tools and documentation practices to facilitate sensemaking and subsequent data reuse. \newline

The key contributions of this work are identifying: (i) patterns of data-centric sensemaking activities; (ii) a framework for these activity patterns and their related data attributes; (iii) user needs for data reuse; and (iv) a set of design recommendations to support data-centric sensemaking.

\section{Background}
\label{background}

Sensemaking has been studied across a range of disciplines, including psychology (e.g. \citet{DBLP:journals/expert/KleinMH06}), decision making  (e.g. \citep{klein1993decision,malakis2013sensemaking}), organizational behaviour (e.g. see \citet{maitlis2014sensemaking} for a review), information seeking  \citep{dervin1997given,DBLP:journals/ijhci/MarchioniniW07}, and human computer interaction (HCI) (e.g. \citep{DBLP:conf/chi/RussellSPC93}). In this work, we focus on sensemaking as discussed in information science and HCI. In these domains, sensemaking is defined as the process of constructing meaning from information \citep{blandford2010interacting}, and is recognised as being an iterative process that involves linking different pieces of information into a single conceptual representation \citep{hearst2009search,DBLP:conf/hvei/Russell03}. \newline

\subsection{Sensemaking and information seeking}
\label{sec:infoseeking}
Models of information seeking behaviour often present sensemaking as a key component. Traditional models detail the specific steps involved in sensemaking during information seeking as a sequential, yet evolving, process \citep{DBLP:journals/iwc/SutcliffeE98,hearst2009search,kuhlthau2004seeking}. While traditional models tend to be static, many of their authors emphasise that people's behaviour is complex and changes when being presented with new information. More recent, dynamic models acknowledge a variety of influencing factors in finding and making sense of information, e.g. skills, knowledge, preconceptions, culture or motivation (\citep{kelly2009methods,klein2007data}). Other work examines the cognitive mechanisms involved, framing sensemaking as a series of different information processing components taking data as input and producing conceptual changes as an output \citep{doi:10.1086/594540,doi:10.1002/asi.24221}. \newline

\subsection{Data-centric sensemaking}
While sensemaking of textual information has been well-explored, there is a relative gap in research that aims to understand the strategies involved in making sense of data. Compounding this is the fact that the very definition of ``data", particularly ``research data" has itself been the subject of much debate. An increasingly common conceptualisation of research data is that proposed by \citet{borgman2015big}: data are representations of observations, objects, or other entities that are used as evidence for the purposes of research or scholarship. This definition does not distinguish between data formats or qualitative or quantitative data, recognizing that what serves as data in one situation for one individual may not act as data in another situation for another individual (see also \citep{Pasquetto2017}). Similary, in their data frame theory, \citet{klein2007data} emphasise how the perspective (or frame) of the data consumer shapes the data in terms of how they are perceived, interpreted and even acquired. Through engaging with data, preexisting frames either change or get reinforced, which can be seen as an aspect of sensemaking. Critical data studies also describe this as a collective process, due to interpretative layers built into the creation and use of data \citep{DBLP:journals/bigdata/NeffTFO17} \newline 

Studies in HCI tend to focus on quantitative data, addressing, e.g., the role that visualization plays in identifying patterns in data \citep{furnas2005making, DBLP:journals/tvcg/KangS12}; this focus reflects the emergence of bespoke visual exploration environments \citep{DBLP:journals/tvcg/YalcinEB18,marchionini2005accessing}. 
Other work proposes tools to aid in sensemaking activities, such as a visual analytics system tailored for particular groups of data analysts \citep{stasko2008jigsaw} or agile display mechanisms for users accessing government statistics \citep{marchionini2005accessing}. Investigations of exploratory data analysis (EDA) strategies, where new data are explored with a set series of procedures until a high-level story emerges, are also of relevance. Common EDA techniques include performing rough statistical checks and analyses (e.g. calculating descriptive statistics) or looking for general trends or outliers in the data \citep{DBLP:journals/jais/BakerJB09, marchionini2006exploratory}. Many EDA techniques are graphical in nature and are undertaken to help assess the quality of the data. \newline\indent

The first phase of getting to know data, which can involve exploratory data analysis techniques, has been shown to involve a high level of cognitive effort \citep{DBLP:journals/jasis/ZhangS14}.
Existing categories can prompt users to activate related memory content, resulting in converging categorization and verbalization processes; this influences how information is interpreted and potentially eases sensemaking efforts \citep{DBLP:journals/chb/LeyS15,DBLP:journals/chb/FioreCO03}.

Engagement and sensemaking with data is also determined by the purpose of the engagement activity, usually connected to a task, which can range in specificity. The importance of quality indicators and uncertainty attached to data is task dependent \citep{Boukhelifa:2017:DWC:3025453.3025738,DBLP:conf/chi/KoestenKTS17}.
While there are a variety of task classifications in the information seeking literature (e.g. \citep{DBLP:journals/ipm/Freund13,DBLP:journals/ipm/LiB08}, to this date there is no established taxonomy for data-centric work tasks, which might reflect rapidly changing work practices with data. \newline \indent

\subsection{Evaluating data for reuse}
There is a growing amount of literature, particularly within information science, that examines the reuse of research data. Key studies question and explore the definitions and types of data reuse within and across disciplines \citep{Pasquetto2019Uses, vandeSandt:2690554}. Many studies characterize the contextual information required to make decisions about using (or not using) data within particular disciplinary fields (i.e. \citep{faniel2017practices,kriesberg2013role}. Although a set definition of context remains challenging \citep{faniel2019context}, there is an overall agreement that data reuse without any contextual reference is almost impossible to do well \citep{DBLP:conf/group/BirnholtzB03,borgman2015big}.

Building on studies of researchers in three disciplinary domains, \citet{faniel2019context} propose a typology of the information needed to support data evaluation, finding that information about data production, data repositories, and data usage are key in making decisions about reusing data. Similarly, \citet{Gregory2020Lost} find that researchers across disciplines rely on information about data collection conditions, data processing, topic relevance and accessibility when evaluating data. This aligns to a large degree with \citeauthor{koesten2019everything}'s findings on dataset-specific selection criteria covering different aspects of relevance, quality and usability \citep{koesten2019everything}. 

Other work looks specifically at how researchers develop trust in data. \citet{DBLP:journals/jasis/Yoon17}, for instance, draws on interviews with quantitative social scientists to explore the social, multi-stage processes involved in trust development. She identifies data characteristics which can aid in building trust, such as the quality of documentation and the reputation of the data publisher. \citet{DBLP:journals/pacmhci/PassiJ18} describe perceived trustworthiness of data or a data science system as a task dependent and collaborative accomplishment that involves assessing different types of uncertainties. While the criteria used for both data evaluation and trust building likely play important roles in data-centric sensemaking, \citet{DBLP:conf/group/BirnholtzB03} highlight that much of the knowledge needed to make sense of data is tacit and not included in data documentation.

\subsection{Summarisation as a way to understand sensemaking}
We adopt a study design that builds on work using summarisation as a way of exploring cognitive processes \citep{hidi1986producing}. Summarisation tasks, as studied in psychology, are described as involving three distinct cognitive activities: selection of which aspects should be included in the summary; condensation of source material to higher-level ideas or more specific lower-level concepts; and transformation by integrating and combining ideas from the source \citep{hidi1986producing}. As comprehension is viewed as a prerequisite for summarisation, text summarisation tasks have been used to assess recall and language abilities \citep{kintsch1978toward}. We build on these ideas and use summarization as a way of exploring the cognitive processes involved in comprehending data as an information source. 

In a recent study, \citet{koesten2019everything} had participants produce written dataset summaries in order to better understand selection criteria for datasets. While these summaries provide insights into the conceptualisation of datasets, written summaries do not always capture the complex verbal sensemaking that precedes their creation \citep{mayernik2011metadata}. Verbal, or spoken, summarizations often reflect deeper, more spontaneous cognitive processes \citep{DBLP:journals/jasis/CrestaniD06}, but they have yet to be used to understand data sensemaking behaviours.\newline \indent

\subsection{Summary of key points}

We argue that in order to reuse data, data consumers must first be able to understand and make sense of those data. While we hypothesise that these sensemaking processes will include attributes similar to those identified in the above literature, we also postulate that data-centric sensemaking involves particular cognitive processes and social and technical interactions resulting in common patterns of sensemaking activities. Our work therefore takes into account not just attributes and categories related to engagement with data, but also considers the wider social, disciplinary and communication contexts existing in data work and their impact on consumer engagement with data.

Our argument builds on the assumption that sensemaking affords specific activities when engaging with data opposed to other information objects (e.g. textual sources), which is mirrored in the literature. However, the sensemaking and information seeking literature often either focuses on textual sources or does not clearly differentiate which source is addressed.

It is also worth noting that there is a significant amount of literature on dataset reuse that focuses on operational problems, machine readability and data interoperability \citep{koesten2019everything}. We do not review this literature in detail here, as our purpose is to focus on the less-studied practices and patterns involved in understanding data. We connect our work to these discussions regarding technical solutions for facilitating data reuse by providing empirical evidence of data-centric sensemaking and by identifying common patterns of sense-making activities to enable design efforts.

\section{Methodology}
\label{Methodology}

Our past work in textual data summarization \citep{koesten2019everything} 
and the reuse of research data \citep{gregory2019understanding} 
informed the the creation of a semi-structured interview design examining how people verbally summarize and make sense of both familiar and unfamiliar data.

\subsection{Study design} 
All participants were asked to bring data that they had used or were familiar with to share during the interview. We refer to this data as the \textit{known data} in this paper. 
We left the decision about what constitutes ``data" up to participants. The majority ($n=27$) chose to bring data which they had created themselves. Most of the data brought by participants were spreadsheets ($n=19$); other data included textual data (e.g. interview transcripts), images, videos or other artefacts. We did not ask for any documentation, supporting information or metadata from participants to see what they brought when not prompted.

\begin{figure}[H]
\centering
\includegraphics[width=125mm]{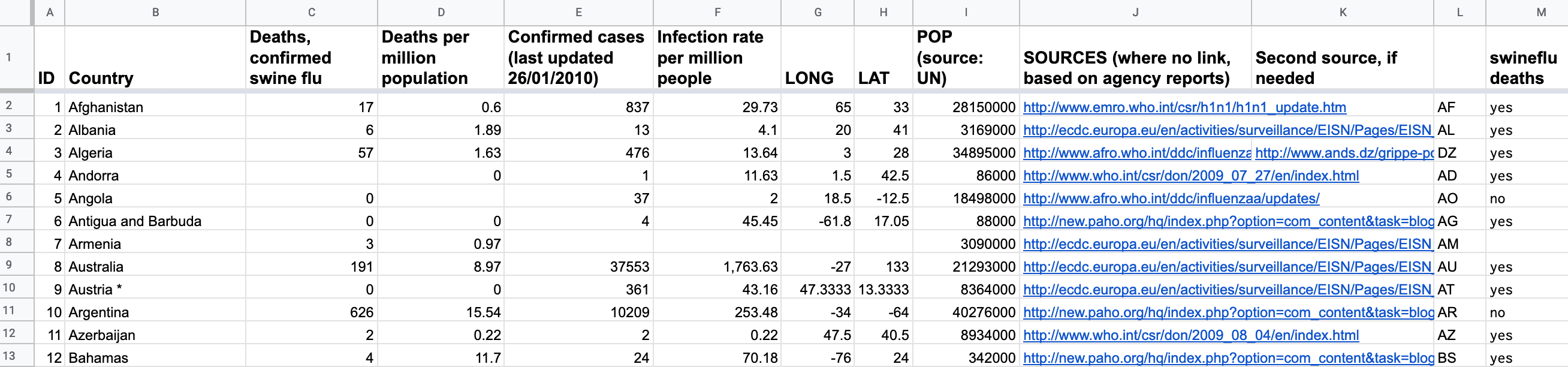}
\caption{Excerpt of the provided or \textit{``unknown"} dataset describing the global occurrence and mortality rate of swine flu}
\label{fig:spreadsheet_1}
\end{figure}

We also prepared a dataset to share with participants; we refer to this data as the \textit{unknown data} in this paper. This dataset was a modified version of a spreadsheet from a popular news source in the UK, the Guardian Data Blog\renewcommand*{\thefootnote}{\arabic{footnote}}, United Nations\footnote{\url{ https://www.theguardian.com/}} which was used in a previous study \citep{koesten2019everything} (see Figure \ref{fig:spreadsheet_1}; the entire spreadsheet is available on a GitHub repository\footnote{\url{https://github.com/laurakoesten/talkingdatasets}} associated with this work). This dataset met specific selection criteria: it included numerical and textual data, missing values, inconsistencies in formatting and some ambiguous variables. At the same time, the data were understandable and not specific to a particular domain.

\subsection{Data collection}
The interview protocol (available on GitHub) consisted of two primary sections: questions about the \textit{known} data, and questions about the \textit{unknown data}. 
Interviews lasted $30-60$ minutes and were held using the web-conferencing application Zoom. All interviews were audio-recorded; screen-recordings capturing participants' interactions with the data were obtained for $26$ interviews. (Five recordings were not created due to technical problems).  
\newline
Both sections of the interview began with the verbal summarization task. The task was for participants to provide a general description of the data to a data reuser who is trying to decide whether to use the data for a particular task, but is unable to see the data. We formulated the task in this fashion in order to elicit rich descriptions of the data, not aimed at a particular use case. 

Further questions were then asked regarding participants' data, their data creation and documentation practices, and data reuse. In the second section of the interview, we shared our dataset and asked participants to perform the same verbal summarization task for the \textit{known data}. We also asked them to describe and discuss specific areas of the dataset and posed follow-up questions about data reuse, data sharing and data search (examples can be seen in Table \ref{tab:interviewschedule}. 

\setlength{\textfloatsep}{0.6cm}
\renewcommand{\arraystretch}{0.9}
\begin{table}[H]{}
\centering
\small
  \begin{tabular}{p{3.8cm}p{7cm}}  
    {\small \textit{\textbf{Section}}} 
    & {\small \textit{\textbf{EXAMPLES}}} \\
    \midrule
    0: Background & Demographics, job role, discipline or research area,  experience of working with data \\
    1: Summary of known data & \textit{Verbal summarisation task}  \\
    1: Context of known data & Information structures needed for reuse \newline Describe for colleagues vs someone outside your domain   \\
    2: Summary of unknown data & \textit{Verbal summarisation task}    \\
    2: Context of unknown data & Anything missing that you would like to know about the dataset  \\
    3: Specific areas of unknown data & Rows / columns with missing / ambiguous data,  different variable types  \\
\end{tabular}
  \caption{Overview of interview schedule including the summarisation task and example topics for the different interview sections}~\label{tab:interviewschedule}
\end{table}

Two pilot interviews were conducted in October 2018. This allowed us to determine the interview duration, to fine-tune the interview questions and the set-up of the summarization task. The remaining interviews were held between November 2018 and January 2019 and were transcribed by a professional transcription firm. \\

\textit{Recruitment.}\\
Our primary sample was drawn from a pool of individuals, past respondents to a large scale survey study conducted by \citet{Gregory2020Lost}, who had published at least one article indexed in Elsevier's Scopus literature database\footnote{\url{https://www.scopus.com}} in the last three years.

We sent recruitment emails ($n=1000$) in November-December 2018 in two batches and received $47$ positive responses. From those, we selected $27$ participants who represented a range of disciplines and nationalities and were proficient in English. We recruited an additional four participants via convenience and purposive sampling, for a total of $31$ participants. Participants for our pilot interviews were identified using purposive sampling. We did not offer incentives for participation in this study. 

\textit{Participants.}\\
Participants ranged from age $26$ to age $73$, with the majority being between $30$ and $45$  years old (Median $40.6$). They reported $19$ different countries of residence worldwide, with a skew towards the Netherlands ($n=5$) and the UK/USA ($n=3$). $13$ out of the $19$ countries are in Europe; $20$ of our $31$ participants live in European countries. 

Although participants work in multiple countries, the majority were fluent in English; minor problems with language or internet connectivity were experienced in two of the interviews. Over half of the participants ($n=18$) worked at a university or college at the time of the study, with six working in research institutions. Participants' disciplinary domains and roles are described in Table \ref{tab:participants}. All participants have previously published research papers. The majority were experienced with quantitative research; 
others categorised themselves as predominantly qualitative researchers, or used both quantitative and qualitative methods.

\setlength{\textfloatsep}{0.5cm}
\renewcommand{\arraystretch}{0.4}
\begin{table}[H]{}
\centering
\small
  \begin{tabular}{p{0.2cm}p{5.2cm}p{4.5cm}}  
    {\small\textit{P}}
    & {\small \textit{\textbf{Domain}}} 
    & {\small \textit{\textbf{Role}}} \\
    \midrule
    $1$ & Biological sciences & Project manager    \\
    $2$ & Life sciences, Paleontology & Project acquisition manager      \\
    $3$ & Biblical studies, Information Technology & Researcher    \\
    $4$ & Musicology, Humanities & Project leader, project manager     \\
    $5$ & Geophysics  & Data curator    \\
    $6$ & Physics, Chemistry & Post doctoral associate    \\
    $7$ & Analytical Chemistry & Researcher     \\
    $8$ & Material Science and Engineering & Professor emeritus, researcher     \\
    $9$ & Social Science (Social Care, Social Work) & Senior research fellow    \\
    $10$ & Social sciences, Computer science & Director of Research Services    \\
    $11$ & Social justice, Socioeconomic Justice &Professor     \\
    $12$ & Geology, Earth Sciences & Research scientist    \\
    $13$ & Earth Sciences & PhD student    \\
    $14$ & Fluid Mechanics & Researcher   \\
    $15$ & Molecular Biology & Researcher   \\
    $16$ & Tourism, Social Psychology & Senior lecturer     \\
    $17$ & Mathematical education & Assistant professor    \\
    $18$ & Telecommunications, Computer science & Associate professor    \\
    $19$ & Biological anthropology & Postdoctoral research fellow    \\
    $20$ & Medicine, Biomedicine & Researcher and teacher     \\
    $21$ & Agriculture, Food science & PhD Student     \\
    $22$ & Medicine & Surgeon, PhD student     \\
    $23$ & Entomology (Biological Sciences) & Researcher, curator    \\
    $24$ & Environmental sciences, agriculture & Lecturer     \\
    $25$ & Biostatistics, Epidemiology & Associate professor, biostatistician    \\
    $26$ & Material Science & Researcher     \\
    $27$ & Psychology & Researcher, PhD student   \\
    $28$ & Veterinarian, Obstetric Clinician  & Assistant professor    \\
    $29$ & Information science, Medicine & Associate director    \\
    $30$ & Environmental sciences &  Researcher   \\
    $31$ & Medicine, Mental Health & Head of research group in a hospital   \\
\end{tabular}
  \caption{Description of participants (P) with their disciplinary domains and professional roles}~\label{tab:participants}
\end{table}

\subsection{Data analysis}
The coding strategy for thematic analysis was developed through a multi-step process of independent parallel coding \citep{thomas2006general}, using the
the qualitative data analysis program NVivo. Two authors independently analyzed a sample of seven interview transcripts and developed an initial codebook with supporting examples, employing a combination of deductive and inductive thematic analysis \citep{robson2016real}. Codes developed through deductive analysis were oriented on the different sections of the interview protocol and on existing literature in data summarization \citep{koesten2019everything} and data reuse \citep{faniel2019context,faniel2017practices}. 

Within these high-level themes, the authors iteratively developed codes based on a general inductive approach \citep{thomas2006general} through sequential readings of the transcripts. 
The independently-developed codebooks were compared for similarities and differences and combined and modified to create a single unified codebook which was then used to re-code the sample transcripts. To further enhance the reliability of the coding scheme, two senior researchers checked and discussed the unified codebook for a sample of the data. \newline

\begin{figure}[H]
\centering
\includegraphics[width=110mm]{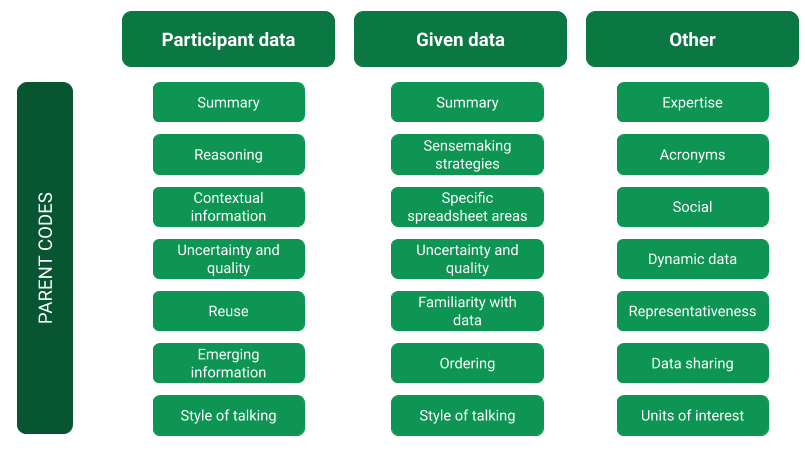}
\caption{Primary codes}
\label{fig:codebook_treemap}
\end{figure}

Based on this analysis, we made further modifications, resulting in a nested coding tree consisting of three primary codes with a total of $30$ child codes (see Figure \ref{fig:codebook_treemap} for the most used codes). We consolidated these codes through axial coding \citep{straus1990basics} drawing out those links which allowed us to answer our research questions. The themes identified through axial coding are used to structure the Findings section and form the basis of the synthesis presented in Figure \ref{fig:activitycluster}.

\textit{Screen recording analysis}
We analyzed the 26 captured screen recordings to identify common interactions with the \textit{unknown} dataset. We examined participants' actions during the general summarisation task. 
Two of the authors independently viewed a sample of these screen recordings to identify common interactions. The authors then discussed the list of interactions and developed a list of $15$ interactions to use in the video analysis. We used this list to identify the first occurrence of each possible action; we did not consider the duration of each action in our analysis. Via the screen recordings, we could document how much of the spreadsheet participants could visibly access on their screens without scrolling. This allowed us to control for larger screens.

\textit{Visual analysis}
All plots were created using the statistical analysis program R. We used the color palette "viridis", as it has been shown to be more accessible than other comparable color schemes.\footnote{\url{ https://cran.r-project.org/web/packages/viridis/vignettes/intro-to-viridis.html}}

\textit{Ethics}
The study was approved by the University of Southampton's Ethical Advisory Committee under ERGO Number 45874. Informed written consent was given by the participants prior to the interview.

\section{Findings}
\label{Findings}
We present our results along two dimensions: the research questions identified in the introduction section and the clusters of sensemaking activities which we identified via axial coding, namely \textit{inspecting} the data, \textit{engaging} with the content and \textit{placing} data in broader contexts. Although we divide this section by research question to improve readability, the evidence we present often spans these divisions. 

We pay special attention to both \textit{activity patterns}, which we define as common physical and cognitive actions undertaken by participants when engaging with the data, and data \textit{attributes}, or characteristics of the data with which participants interacted. We examine the findings in light of data reuse and synthesise them in the Discussion section to provide an overview of the patterns we identify.
\subsection{RQ1: What are common patterns in sensemaking activities, both for known and for unknown data, in the initial phases of data-centric sensemaking?}

\subsubsection{Inspecting}
\begin{figure}[H]
\centering
\includegraphics[width=100mm]{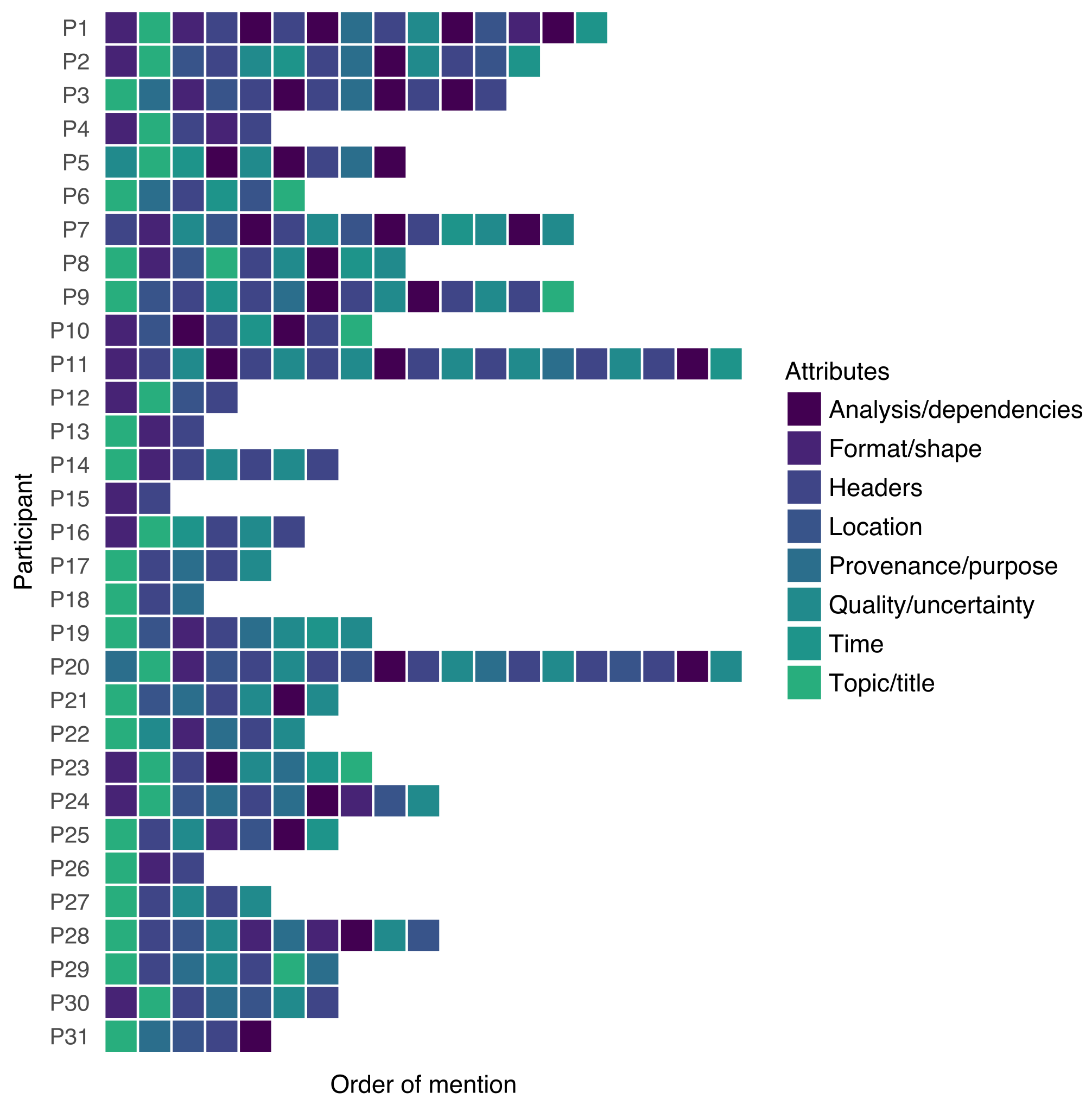}
\caption{The order in which participants discussed certain attributes in the \textit{unknown} dataset}
\label{fig:process_diagram_givendata}
\end{figure}

\begin{figure}[H]
\centering
\includegraphics[width=100mm]{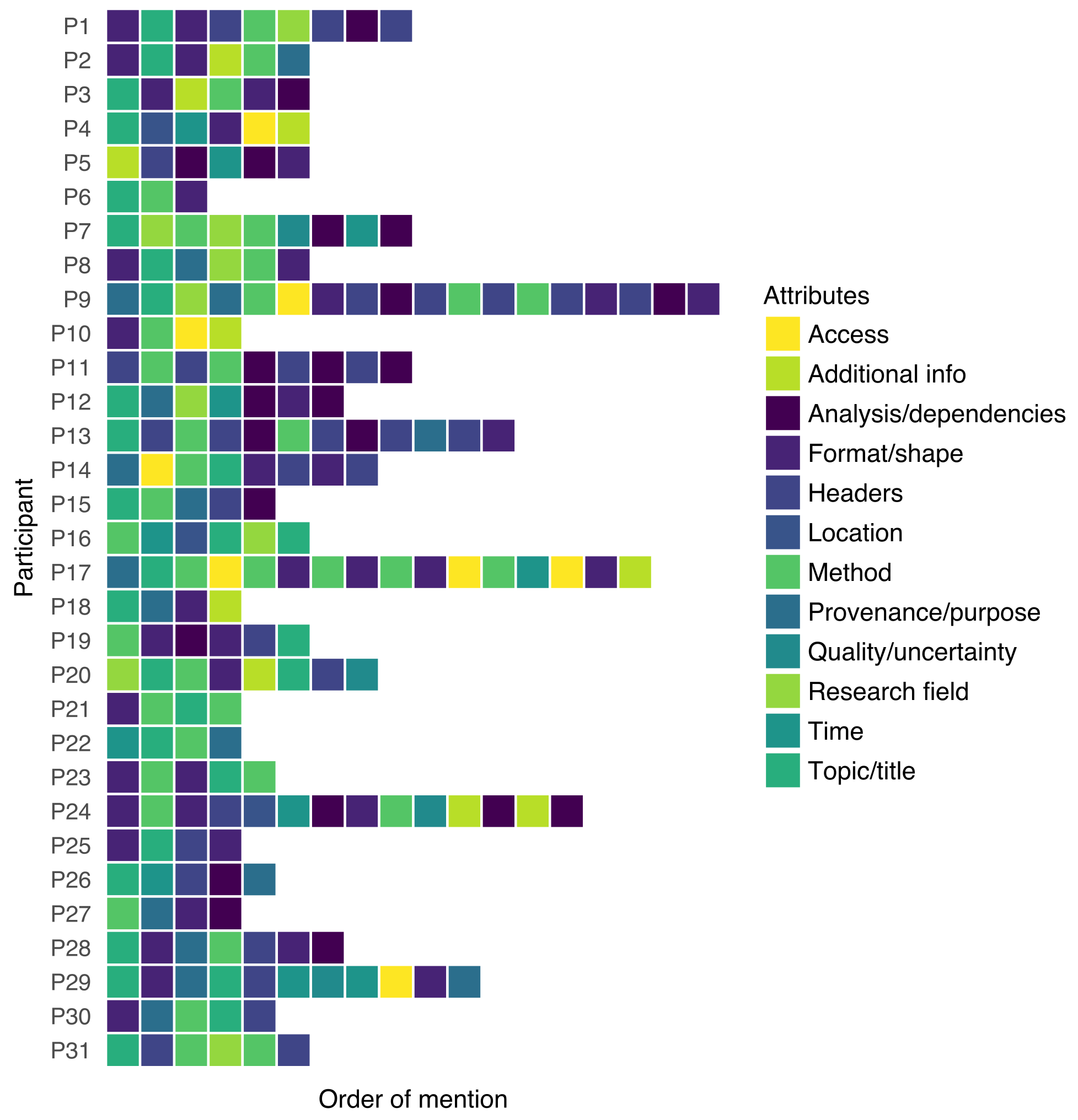}
\caption{The order in which participants discussed certain attributes in their own data (\textit{the known data}).}
\label{fig:process_diagram_theirdata}
\end{figure}

\begin{figure}[H]
\centering
\includegraphics[width=100mm]{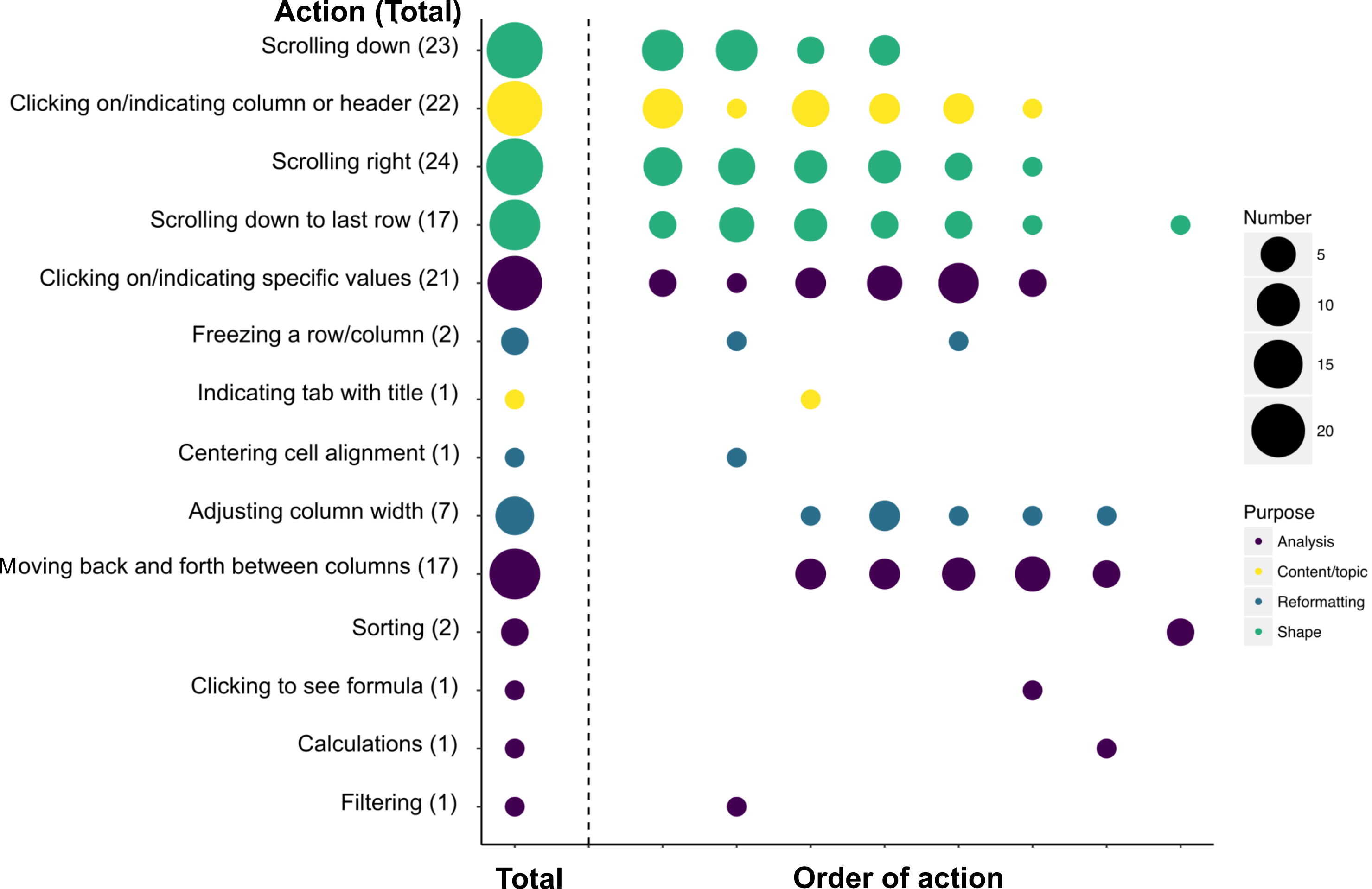}
\caption{Total number of actions and order of actions observed in screen recordings of the \textit{unknown} dataset. Size of circle represents number of participants engaging in activity. Figure is arranged according to which activity most frequently occurred first. Color represents purpose of action.}
\label{fig:video_plot}
\end{figure}

When participants were first shown our dataset, we asked them to perform the verbal summarization task -- to provide a general description, after taking a few minutes to explore the data silently. In this section, we examine both the order of how participants discussed attributes of the data (see Figure \ref{fig:process_diagram_givendata} and  \ref{fig:process_diagram_theirdata}) and their actions in the spreadsheet during these verbal summarisations (Figure \ref{fig:video_plot}).

\textit{Order of verbal summarisation} 
We observed two approaches when completing the verbal summarisation task: participants took either a linear or an interwoven approach. In linear summaries of the \textit{unknown} data ($n=23$), participants addressed the data attributes identified in Figure \ref{fig:process_diagram_givendata} (e.g. time, location, format) one-by-one before proceeding to the next attribute. In the interwoven summaries, participants interspersed descriptions of individual attributes with analyses and comments ($n=8$). 

Figure \ref{fig:process_diagram_givendata} shows the attributes headers ($n=64$), quality/uncertainty ($n=42$), topic/title ($n=33$), and analysis/dependencies ($n=30$) were most frequently mentioned in descriptions of the \textit{unknown} dataset. The majority of participants mentioned the overall topic or title as one of the first two attributes ($n=24$); roughly half of participants mentioned the format or shape of the data (e.g. the number of columns, rows or observations) either first or second ($n=15$). The discussions of other attributes were likely influenced by the structure of the dataset itself. Location information was a prominent part of the dataset, e.g., as the data were ordered by country and the four columns containing geographic information were positioned on both the left and right sides of the spreadsheet. The data included only minimal temporal information. The majority of general descriptions mentioned location ($n=22$) toward either the beginning or end of the description, while temporal information was mentioned in just under half of the general descriptions ($n=13$).

In the linear general descriptions, time and location were discussed or questioned at a general level:

\begin{quote}
This communication shows us the deaths from swine flu in the countries around the world, Afghanistan, Albania, Columbia, Bolivia. (P31)    

The one thing that is not apparent immediately from the data is the time span. (P19)
\end{quote}

Participants taking an interwoven approach to summarization engaged in more initial analysis, repeatedly seeking relationships and dependencies between the spreadsheet columns or expressing uncertainty about meaning or the quality of the dataset.

\begin{quote}
I don't see any date or year, for purposes of comparison then it's a bit problematic, I can for example only do comparison charts for those with an asterisk for Austria and Bulgaria, for example, because they all have the data from 2009 but for number of deaths recorded in that country, then this data is useful, infection rate per population. (P5)
\end{quote}

We observed similar attributes within the general descriptions of participants' own data, but participants also mentioned additional attributes, i.e. details of their own field of research, methodology and details of the particular study, data availability and access restrictions, and the existence of additional information or documents needed to describe and understand their data (Figure \ref{fig:process_diagram_theirdata}).

Most of the general descriptions of participants' data followed a linear pattern ($n=24$). This could be  because participants were not working to understand their own data, but were rather aiming to make their data understood. They were also to some extent better prepared for the requirements of the task, having already had experience working with and discussing their data. In interwoven summaries of their data ($n=7$), participants mixed descriptions of study methodologies with descriptions of headers, variables and data format; some relied on methodological descriptions to communicate the general topic of their data. 

\begin{quote}These are experiments from a 50 metre long indoor set up that we have, where we ran gas and oil through the pipeline, through a 60 metre long pipeline, and we measured the average values - so pressure drop and build-up. And we did that for different gas and liquid velocities, and they also changed the type of oil, so we did this with one oil with a quite low viscosity and one with oil with a quite high viscosity. (P14)
\end{quote}

\textit{Actions in the spreadsheet}
The actions captured in the screen recordings of the verbal summarization task for the \textit{unknown} data support the attributes identified in Figure \ref{fig:process_diagram_givendata} and \ref{fig:process_diagram_theirdata}. Figure \ref{fig:video_plot} shows the total number of actions observed, as well as the frequency of their order of occurrence. Scrolling right ($n=24$) was the most frequently observed action, followed by scrolling down ($n=23$). Participants also clicked on or indicated column headers and specific values. Clicking on both headers ($n=18$)  and particular cells ($n=17$) occurred more often than other forms of indicating these areas of the spreadsheet, i.e. hovering over, or circling them. 

Analysing the order of these actions show that the majority of participants began by determining the length, breadth and general topic of the \textit{unknown} dataset. Nine participants first scrolled down, while eight clicked on headers and seven initially scrolled to the far right of the spreadsheet. Once participants established the general shape of the data, more analysis-related actions were observed, most noticeably examining specific cell values by clicking or indicating and moving back and forth between different columns. One example of left and right scrolling was switching between the different types of geospatial columns which were not located in close proximity to one another. 

Some participants prepared for analysis by reformatting the spreadsheet ($n=10$), i.e. by adjusting the column width or freezing columns or rows. (For three recordings, the width of one of the columns was not optimised to allow reading one of the header names).  Analysis features of the spreadsheet were only used in four instances, for actions such as sorting, filtering, or performing calculations. This reflects the nature of the think aloud task and the time limitations of our study.

We began examining screen recordings for participants' data after the general description task. These screen recordings provide a different type of insight, revealing actions that participants took to ensure that the interviewer adequately understood their data. The actions that we observed ranged in complexity. Participants with spreadsheet data often clicked on each column header, as they provided more detail about each column. Others demonstrated how they would analyse the data, showing unique functions of their analysis software or creating sample spreadsheets and plots. 

\subsubsection{Engaging with the content}
Participants engaged with data content in more depth as they worked to explain and understand the data. This stage of deeper interrogation sometimes began during the scanning phase; it occurred both when interacting with the \textit{unknown} dataset and with the \textit{known} data. Table \ref{table:quotetable}, which is further discussed throughout this section, presents quotations demonstrating similarities and differences in how they engaged with both known and unknown data. This table is organised along three themes: \textit{encodings}, or codes developed to understand and interact with data, \textit{acronyms and abbreviations} and \textit{identifying ``strange things"} within the data.

\renewcommand{\arraystretch}{1.6}
\begin{table}[H]
\centering
\scriptsize
\begin{tabular}{p{1.2cm}p{5.9cm}p{5.9cm}}
 &\textbf{Known Data} & \textbf{Unknown Data}\\
\hline
\textbf{Encodings}& 
(P23): Because of the way it's set up, while it may appear on screen in words, it's actually all in zero, one, two, three, four up to nine. But you can present it yourself in words, and that's really helpful if you're scoring, because you can actually click from one cell to the next, and down the base it will actually tell you what that character is in that cell and whether you just code a zero or a one. 
\newline \newline (P9): Age band and then I grouped the age bands into adult and older people, and that was one of the issues of each of our journeys, a different way of categorising people, so I ended up with a very broad age range really.
 & (P28): I don't know if it could be interesting to be banded in categories like, I don't know, continents... it depends.
 \newline \newline (P19): It looks like they started to code for if there are no deaths, then it's coded as a zero, but there are some instances where there are missing data.\\

\textbf{Acronyms and \mbox{abbreviations}}
& (P22): That is a classic abbreviation in the field of hepatic surgery. AFP is alpha feto protein. It is a marker. It's very well known by everybody...the AFP score is a criterion for liver transplantation. \newline \newline
(P28): So if there is strange code that people cannot understand, I make a legend. Normally the colleagues I'm working with, we use our terms, so I tend to use the most user terms like LD; like SEM is typical for...everyone in my field.
& (P20): If I would make this assumption, I would say this is like geographical location of the countries, but I have no idea what is `Long' and `Lat'. In my work, I have never encountered these kind of acronyms, so it's currently hard for me to assume what would this mean in the context of swine flu. \newline \newline 
(P7): I'm not sure what `long' means. I wonder if it's not something to do with longevity. On the other hand, no, it's got negative numbers. I can't make sense of this. \\
\textbf{Identifiying "strange things"} & (P7):  Let's say from previous experiments and or runs, you know that repeating the experiment, you would get within an error of say 5\% or 2\%, whatever the case maybe. So obviously these three [indicating error bars] are huge, and it would mean that you will have to repeat. So either something is wrong in your system, or you get something wrong during the sample preparation, or the system's not stable, or something else is going on. Or that you're just not planning enough repetitions to get to the true value, so I think it is an important measure to determine if you've got reliable data. & (P20): If I would not go into those cases, like with these discrepancies, I would just assume that this column indicates only the optimised data about whether they're aware or they're not, that [deaths are] due to swine flu in these countries." \newline \newline (P14): That is simply a column saying if there are any deaths at all or not for a certain country related to the swine flu. I see there is a formula here, just simply checking if Column L is larger than zero. So exactly using this information...so then that means there is something wrong with the formula or I completely misunderstood what Column L is.\\
\end{tabular}
\caption{Exemplary quotes illustrating participants' interactions with both their own and the \textit{unknown} data.}
\label{table:quotetable}
\end{table}

In this phase, participants identified patterns and trends (e.g., via simple analyses or discovering relationships between columns) and discussed encodings, often related to categorisations, expressed within the data. They also explored uncertainties attached to the data and the data's overall integrity. We point out two particular instances  observed in this level of engagement with the data: understanding strange things and collaborative sensemaking. \newline

\textit{Data analysis, encodings and tools}

When discussing their data, participants demonstrated how they seek patterns and relationships by creating plots, switching between layers on geospatial images, and developing scales and formulas. Participants also expressed a desire to create plots to visualize the unknown data to identify trends and sought anchor variables as they investigated individual columns and described sample rows. They further drew attention to columns with limited value ranges in their descriptions, e.g. columns with binary variables or those with only a few categorical variables; fewer analyzed the range of values in columns with continuous variables.

Participants ``encode" their own data in ways that help them more easily identify trends and generate findings by, e.g., converting categorical variables to numerical values and vice versa. These encodings are often influenced by the specifications of the analysis tools and software which participants use, such as SPSS, R, or domain-specific programs; which can also influence how participants structure their data, at times increasing the data's machine readability.

\begin{quote}
I use this data to create variables in SPSS. The one I'm looking at now has still got all the labels as words; I thought it would be easier to look at as a spreadsheet. There's another process I went through to translate the words into numbers. For SPSS, you really need numbers in the value labels. That was a whole process, to go through of coding the written, the categories, but just adapting those into numbers that I use. (P9)
\end{quote}

\begin{quote}
[We are] working in R and our supervisor wrote a package which can easily work on it, but the main aspect is that you have to have grouping variables and independent variables which are the sensor signs.  Then you have to separate the data to these different types, so the grouping variables and the independent variables because the PC and the IDA in the R can work in this structure. (P21)
\end{quote}

Other forms of encoding included developing broad categories or groupings to describe and analyze data, such as differentiating between raw data and derived data, or numerical and non-numerical data. Participants also created groups of certain columns according to their semantic meaning; demographic variables were mentioned together, as were descriptive attributes for the same instance, e.g. ``columns with sources" or ``socioeconomic measures". These types of encodings were observed when participants discussed both their own and the \textit{unknown} data. When working with our data, participants also searched for how null values were encoded and represented (see Table \ref{table:quotetable}). 

While the majority of participants reported using spreadsheets or Microsoft Excel at some point in their data workflows, very few actually made use of the built-in analysis tools in our spreadsheet at any time during the task. This could be due to time limitations during the interviews or to the fact that participants were not familiar with the Google Sheets environment which we used. It could also be a result of the fact that some participants do not use spreadsheets to analyze data directly, but rather reported using them for other purposes, such as recording and organizing data or cleaning and preparing data for analysis. Spreadsheets are also used by participants to specifically enable sharing data in a way that is easily accessible or compatible with a variety of analysis programs, facilitating data reuse.  \newline

\textit{Expressing uncertainty, seeking quality and understanding strange things}

Both when discussing their own data as well as when engaging with our data, they expressed concerns about potential misinterpretations, focusing on questions that could arise due to misunderstandings about how data were cleaned and processed. For both quantitative and qualitative data, participants viewed the encodings and categories that they had constructed as major risk points for correctly interpreting their data. The encodings that facilitated their own use of the data (as a data producer) may not be helpful or be explained well enough to enable appropriate data reuse by potential consumers of their data.  

\begin{quote} Although we've tried really hard, because we've put in a coding frame and how we manipulate all the data, I'm sure that there are things in there which we haven't recorded in terms of, well, what exactly does this mean? I hope we've covered it all but I'm sure we haven't. (P10)
\end{quote}
They also questioned and critiqued the meaning of the \textit{known} data, highlighting the lack of contextual information about how the dataset was created and the use of unexplained abbreviations in the dataset. When discussing their own data, however, participants often referred to unexplained acronyms or abbreviations common in their own disciplinary domains (see Table \ref{table:quotetable}). 

Participants combined their interpretations about the meaning of our dataset along with analyses of its completeness and how missing values are reported to make quality determinations. They also used missing values as checkpoints to identify relationships between columns and to identify potential errors or anomalies in the data. 

\begin{quote}So the data is fairly complete with really limited missing values, so the quality of data looks good. (P29)
\end{quote}

\begin{quote}
It's got some blanks, which I presume means no data has been given. Although that's interesting...there's some missing data which shouldn't be missing. Because Armenia, for number seven say, it reports three deaths and yet the swine flu deaths is blank, so that's a bit of an anomaly, and there are quite a few blanks actually. (P9)
\end{quote}

Participants looked for other unexpected values (e.g. outliers) or inconsistencies in formatting or standard ways of reporting to assess the precision and accuracy of the \textit{unknown} dataset. Wrestling with these strange things often served as the entry point to a deeper engagement and understanding of the data, allowing participants to question their assumptions and initial understandings (see Table \ref{table:quotetable}). 

\begin{quote}
Now that sounds quite high for the Falklands. I wouldn't have thought the population was all that great...and yet it's only one confirmed case. Okay [laughs]. So yes...one might need to actually examine that a little bit more carefully, because the population of the Falklands doesn't reach a million, so therefore you end up with this huge number of deaths per million population [laughs], but only one case and one death. (P23)
\end{quote}

\begin{quote}
Some of them have decimals, like a lot of decimals, and some don't have any decimals. So I don't know whether that means that those are supposed to be measured more precisely...or that there is an inconsistency of using the amount of decimals per cell. (P1) 
\end{quote}

Encountering the unexpected in their own data is a critical and normal part of participants' research processes (see Table \ref{table:quotetable}).  While anomalies can be indicative of possible mistakes or points for improvement in the study design, they can also reflect unexpected external changes to the study environment, e.g. people withdrawing from a study, or new technologies that have been adopted over the course of long-running studies. Participants repeatedly emphasized the need to communicate information about these changes or potential sources of error to possible data reusers:

\begin{quote}
When I'm explaining the dataset by sharing a screen or showing them the file or to someone who would probably understand the data, from a dataset perspective, I would basically talk about the implausible values and the missing values and if I'm aware of the issues related to the data, I would like to point them out. (P29)
\end{quote}

\textit{Sensemaking through ``collaborations"}
Working with team members is key to making decisions about study design and analysis, i.e. deciding which data are important to record and analyse, how to develop scales, clarifying study details and making sense of mistakes or unexpected values in the data.

\begin{quote}
I know roughly what it consists of, but I didn't know precisely, and I had to go back to the person who generated it and say ``What does column D mean? And where is the location of the thermocouple whose temperature is measured in column E?". (P8)\end{quote}

\begin{quote} We have a table with...almost 30 columns with variables that were collected, including the names of the people who went into the field and collected each of the samples. So we are keeping track of who's responsible for each of the samples, then if we find any error, any mistake, then we can contact those people. (P24) \end{quote}

During the interviews, participants also collaborated with the interviewer to ensure that the interviewer correctly understood their data. Often, important details crucial to understanding the data emerged only when both the interviewer and the participant could see and interact with the data together. We saw this, e.g., in the case of learning about the importance of temporal information in coral reef imaging data or highlighting a key variable (inflammation) in a study about bipolar individuals. For some, it was nearly impossible to explain their data without being able to indicate specific areas of an image or demonstrating how error analysis was conducted.

\subsubsection{Placing}

\begin{figure}[H]
\centering
\includegraphics[width=55mm]{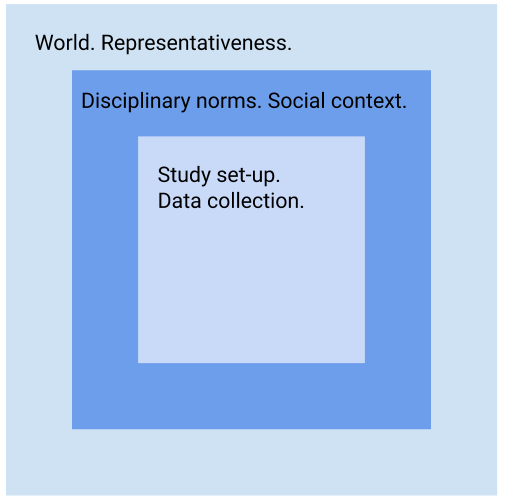}
\caption{Placing data in contexts}
\vspace{-10px}
\label{fig:placing}
\end{figure}

As they engaged more deeply with data, participants placed data into existing contexts, practices and knowledges; this process of ``placing" occurred at different scales (Figure \ref{fig:placing}). Data were placed within their immediate contexts of creation, e.g., when participants detailed study designs, experimental setups or the conditions surrounding data collection, including broader temporal or geographic details. 

\begin{quote}
And it describes, or rather it comprises the results of a laboratory experiment lasting about an hour in which the experimenter,[..], is inducing the crystallisation of a salt in a porous rock sample. And as the crystallisation proceeds, two things happen. One is that the heat is evolved and so the temperature changes and the rock sample slightly increases in temperature, and we measure that temperature at three different positions. In addition there is a very slight expansion of the rock which we detect from the output of a very sensitive mechanical gauge. And then these four measurements, the three temperatures and the mechanical strain are measured at intervals of one second over a period of a few hours and so the dataset consists of this set of numbers. (P8)  
\end{quote}

These types of contextual details have the potential to impact the meaning which the values themselves carry. 

\begin{quote}
Error bars depend a lot on the experimental conditions and on the condition of the material. So, for example, if it was used on powder samples then the error bars would be bigger than the ones that were obtained on single crystal data. (P6)
\end{quote}

At a broader level, participants conceptually placed data within the norms of their disciplinary domains, referencing discipline-specific methodologies and limitations, ways of analysing and verifying data or common data formats. They also recognised that broader social contexts can influence the sensemaking process.

Finally, participants attempted to place data within the world, gauging how representative data are of a particular phenomena. These judgements reflect assumptions about how much the data reflect reality, as data themselves are usually samples which are hardly ever complete, unbiased or without conflict or ambiguity. 

\begin{quote}
It's a pretty large sample size, again, $1,260$. We have equal numbers of males and females. We have three ethnicities: Caucasian, African American and Hispanic, equal numbers of each of those. So it's a well-balanced data set and, because of that, if you were to be interested in how these different cultural values vary or not based on ethnicity, it would be an excellent dataset. (P19)
\end{quote}

One simple example we observed in our dataset was a contention about the representativeness of the countries, which showed a range of interpretations and was expressed in a variety of ways. Participants questioned both the completeness of the list of countries and also whether the data represented the entirety of each country.

\begin{quote}
P2: It's listing the countries for which data are available, not sure if this is truly all countries we know of...

P8: It includes essentially every country in the world

P29: Global data

P30: I would like to know whether it's complete...it says 212 rows representing countries, whether I have data from all countries or only from 25\% or something because then it's not really representative.

P7: If it was the whole country that was affected or not, affecting the northern part, the western, eastern, southern parts 

P24: Was it sampled and then estimated for the whole country? Or is it the exact number of deaths that were got from hospitals and health agencies, for example? So is it a census or is it an estimate?
\end{quote}

During placing activities, participants commonly reported the need to know the original purpose for which the data was created. Descriptions of their own data's original purpose were often complex, as they were intermingled with descriptions of the field of research. Participants floundered in their attempts to place our data, in part because the original study objective was unknown. 

Although important across all dimensions of sensemaking, disciplinary and data expertise were key to placing data. Most participants felt that it was easier to describe their data than to summarise and try to understand our data.

\begin{quote}
My data are much more easier, for sure, because I knew what I was talking about. I didn't have to go through, to understand, which was the quality of the data; I didn't have to understand what it was, the kind of information that this data was giving to me. If I have to go to a database that I've never seen normally and also that is not in my field, it is absolutely much more difficult. (P28)
\end{quote}

\subsection{RQ2: How do patterns of data-centric sensemaking afford potential data reuse?}

Some participants believed that only experts from within their same discipline could reuse their data meaningfully, citing the specificity of their data or the need to analyze the data using specific programs. Others stated that appropriate reuse would require a deep understanding of evolving domain research practices; many had difficulty imagining alternative uses for their data outside of their area of research. 
\begin{quote}

I don't think it would be used for a radically different purpose, but I could imagine somebody taking the data and reanalysing it in relation to a different model of the underlying process, for example.  Or confirming the interpretation that we've placed upon the data using our own model...But they would be people who'd be very close to the topic. (P8)
\end{quote}

\subsubsection{Structures needed for sensemaking and reuse}
A few participants believed that the use of common data structures, terminologies and methodologies within their domains made it possible for their data to ``speak for themselves" to others with similar expertise. 
\begin{quote}
I probably wouldn't have to describe it [the data].  Probably they would just get it. (P1) 
\end{quote}

We observed procedural reasons why data do not speak for themselves, but require additional structures to convey meaning. Participants did not always include column names in their data in order to make them more machine readable, e.g., or they divided datasets into various sub-sheets to ease processing. We also observed that additional information structures,  i.e. documents and codebooks, are needed to support reuse for data consumers regardless of domain, as well as to support future (re)use by the original data producer. 

\begin{quote}Ten years makes a big difference in my memory, too. So, even at the time when I was working in it, I didn't have to refer to the code book, I knew it all by heart. I would have to go back and look at the code book now myself, and that's why it's important to keep the notes on what you're doing with the variables and keep a copy of the survey that was used, the research instrument, those sorts of things. (P11)
\end{quote}

Participants described a large variety of documentation and knowledge transfer practices surrounding datasets (Table \ref{tab:supplemental_things}). These practices and the formats used to provide additional information are shaped by journal restrictions, metadata schemas and repository requirements, and by the perceived usability of the information structures themselves.  Sometimes this additional information is separate from the data; other times it is embedded within the data, i.e. in the case of annotations or descriptions of codes within a spreadsheet. Different data consumers may require different information structures for the same data. 

\begin{quote}
If they're using a different program, I can direct them to a character set, which you can get from this matrix, but the publication of that character set is quite separate but available online. (P23)
\end{quote}
\begin{quote}
So if you start with the README here, then we can take several directions. So, you can delve into the features, what they mean, and you can delve into the feature documentation. You can delve into ways to query it, and do that for yourself, and then you go to all kinds of programming documentation.  And then, here I also pointed to tutorials, [..].  And you can read some papers about it and they're also cited...We also have a Slack community with 120 people, and if they have really hard questions, we invite them to Slack, and they are being answered by either me or people who know more about it. (P3)
\end{quote}

\setlength{\textfloatsep}{0.1cm}
\renewcommand{\arraystretch}{0.9}
\begin{table}[H]
\centering
\small
  \begin{tabular}{p{10cm}}  
    {\small\textit{INFORMATION STRUCTURES}}\\
    \midrule
    Supplementary files (corresponding spreadsheets, text documents, README files) \\
    Resource description document (including, e.g., explanations of columns)  \\
    Code     \\
    Documentation of the code    \\
    Emails /  communication protocols \\
    Figures, visualisations     \\
    Code book / sheet; can contain personal data \\
    Repository     \\
    Presentations / slides     \\
    Technical reports     \\
    Publications     \\
    Maps     \\
    Audio folder     \\
    Slack channel    \\
    Annotations \& interpretations  (also on various levels of the data, e.g. on image layers)  \\
    Tutorials     \\
    Questionnaires / surveys  (variables often created in order of the questions)   \\
\end{tabular}
\vspace{5px}
  \caption{Information structures supporting sensemaking}~\label{tab:supplemental_things}
\end{table}

The study also revealed attributes which should be present in information structures to avoid losing meaning and to enable data reuse. We present these attributes according to two perspectives which emerged in the interviews: the data consumer's distance to the data and the methodological approach of the original study in Table \ref{table:distancetothedata} and \ref{table:methodologicalapproach}. We define ``distance to the data" in terms of a data consumer's familiarity and expertise with particular data. Someone ``close to the data" will have more knowledge of the data and how they were created; 
someone more distant from the data will not have this knowledge. In Table \ref{table:methodologicalapproach}, we focus on two broad approaches to data collection: quantitative and qualitative methodologies. These tables do not aim to present a comprehensive list, but rather reflect the specific work scenarios of our participants. \newline

We asked participants if they would describe their data differently to a colleague with similar expertise, i.e. someone close to the data. 
Rather than needing less information about the data due to prior knowledge, many participants believe that individuals with similar expertise need more granular information about data creation conditions, prior work which the data builds on, and the potential uses of individual variables. Some participants said that they would not describe their data differently to someone close to the data, emphasising instead common attributes that would be important, regardless of a data consumer's distance to the data (Table \ref{table:distancetothedata}). 
\renewcommand{\arraystretch}{1.1}
\begin{table}[H]
  \begin{center}
    \footnotesize
    \setlength{\extrarowheight}{1.5pt}
    \begin{tabular}{p{4.2cm}|p{3.6cm}|p{3.2cm}}
     \toprule
       {\textbf{Close to the data}} & {\textbf{Far from the data}}& {\textbf{No difference in distance to the data}} \\ \midrule
        More granular information about \mbox{conditions}, assumptions, errors, trends, possible questions the data can answer, variable types, analysis \mbox{/ programming} details, sample creation details, study objective & More granular information about research \mbox{explanation}, explanation of all \mbox{abbreviations} / acronyms, how ratios / errors / columns derived & Supplemental materials\\ 
        Benefits / problems of data & Less technical language & \mbox{Study objective and} \mbox{expected} outcome\\
        Previous work that this data builds on, relation to standards in discipline, out-of-discipline abbreviation  &  Research explanation  & Data collection details\\
        Less granular information about field of research, common abbreviations, data format / structure & Tailor to field of interest of the data consumer   & Sample details\\
        & More general data presentation & Potential use of data \\
        & Calculation of ratios/ standard deviation & \mbox{Usage restrictions}, \mbox{confidentiality} concerns \\
        && \mbox{Explanations of codes}, \mbox{categories}, scores\\
      \bottomrule
     \end{tabular}
    \caption{\textbf{PERSPECTIVE: Information needs related to distance to the data.} Someone "close to the data" will have more knowledge of the data and how they were created. Someone "far from the data" will not have this knowledge.} 
    \label{table:distancetothedata}
  \end{center}
\end{table}

\begin{quote}
So if I'm talking to somebody who is data agnostic or who has not worked in a data science field, my description would be limited to the basic variables, the fields that are of interest to the person...If I'm talking to a data science person or a data scientist who's going to use the data, my description would be more granular. My description would be more helping the person understand the benefits as well as the problems associated with the data. (P29)
\end{quote}

\begin{quote}
I would maybe shorten up some things and focus on some others.  For instance, I would expect that everyone I'm working with expects to code BMI in kilograms and to have birth weight in grams because it's a standard unit for those things in Danish health research...I would tell them more about the study design, because often people I work with are epidemiologists. So there one of the main things would be, where do these 2,000 women come from? Is it data from Denmark or from somewhere else?  Is this from last year or from 30 years ago? Things like that, so more complex information so that they can decide if it's relevant for their interests. (P25)
\end{quote}

Different methodological approaches also elicited particular details, although these details were not mutually exclusive of each other (Table \ref{table:methodologicalapproach}). For quantitative data, participants reported needing extensive information about an experimental setup, including how experimental designs differed from the real world environment. 

\begin{quote}
Well I would perhaps mention the size of the pipe diameter. That is something that they're often interested in, because in real pipelines, the pipe diameter is perhaps 12 inch and more, quite large, while in typical labs, you don't have this possibility. (P14)
\end{quote}

\renewcommand{\arraystretch}{1.1}
\begin{table}[H]
  \begin{center}
    \footnotesize
    \setlength{\extrarowheight}{1.5pt}
    \begin{tabular}{p{5.7cm}|p{5.5cm}}
     \toprule
       {\textbf{Quantitative}} & {\textbf{Qualitative}} \\ \midrule
        \textit{Detailed experimental set-up:} \newline including time period, instrument settings, \mbox{location}, etc.; where the test conditions differ from real world settings or from standard procedures & \textit{Detailed study set-up:} \newline including time period, description of participant sample, sample size, mode of interaction (e.g., \mbox{online} or in person) \\
        Who did which work (data collection, quality control, data cleaning, code, analysis) & Who did the research; researcher's relationship to participants and how this was mitigated (e.g., professional role of researcher / 
        context of \mbox{recruitment)} \\
        Are measurements individual measurements or multiple measurements of the same thing that were aggregated & Questions or schedule for surveys or interviews (including information about answer modes (e.g., \mbox{predefined} answers or free text)\\
        Which section of the object was measured, on how much material a measurement was made  & Analysis (e.g., type, coding strategy, groupings and narrative of categorisation) \\
        Standard error, precision of measurements, uncertainties & How sample was chosen (inclusion / exclusion criteria), created, scope and characteristics of sample (e.g., age of \mbox{participants})\\
        Influencing factors (seasonal differences, \mbox{external} events, etc.) & How categories were chosen, how scores were created, variables of focus \\
        Standard units of measurement in a field / setting of study (e.g., instruments -- specifications, reliability, how calibrated, how they work and how they create the data output, software format used to capture or analyze data) & Social context \\
        Number of repetitions of experiment  &   Description of labels / codebook / account of \mbox{variables} \\
      \bottomrule
     \end{tabular}
    \caption{\textbf{PERSPECTIVE: Methodological narrative}  \emph{(characteristics are not necessarily unique to either approach)}} 
    \label{table:methodologicalapproach}
  \end{center}
\end{table}

Key findings from the qualitative perspective include the choice of categories, questions of representativeness and details of the study set-up that influence the data, such as whether participants are required to answer a survey question. Social context also influences how study participants communicate, e.g. in the case of interview participants in conflict areas who may not feel safe enough to respond truthfully to questions.

\section{Discussion}
\label{Discussion}

\begin{figure}[H]
\centering
\includegraphics[width=150mm]{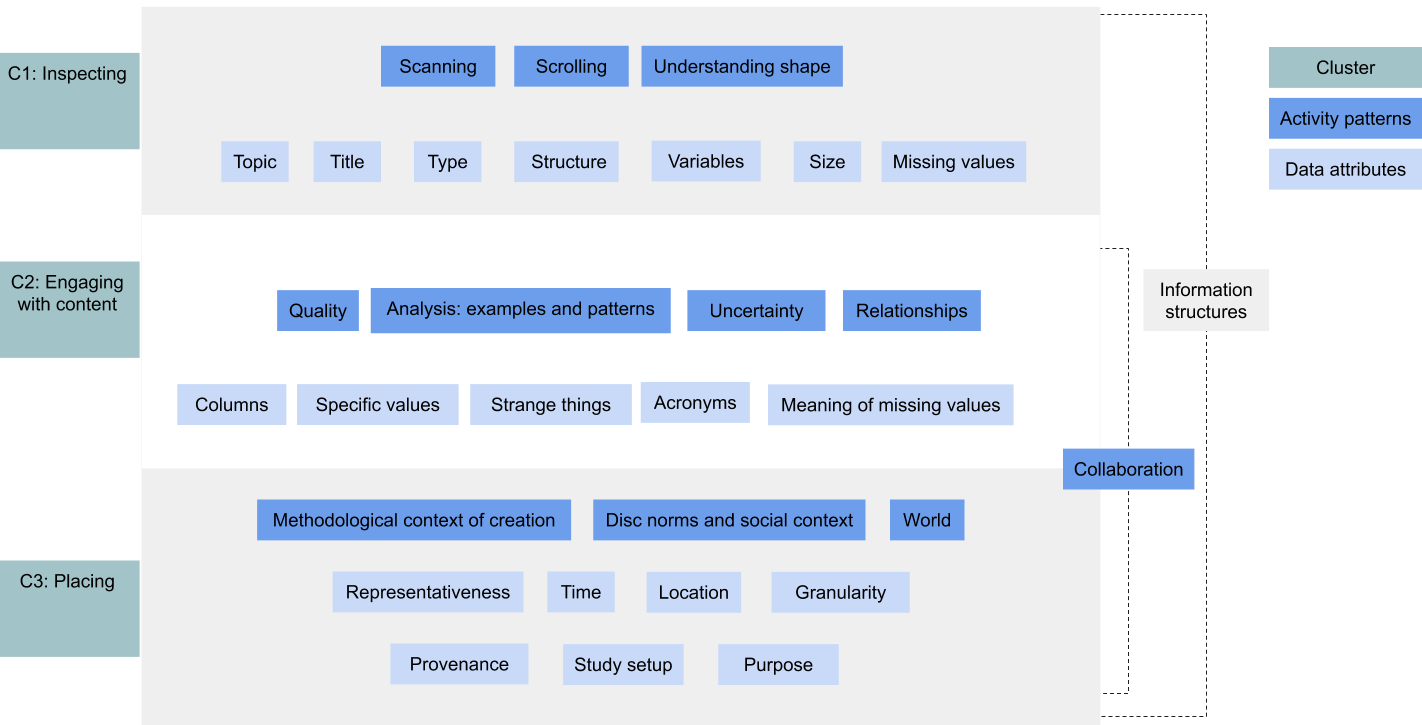}
\caption{Activity patterns and attributes in data-centric sensemaking}
\label{fig:activitycluster}
\end{figure}

We bring together different perspectives in this study, drawing together our findings about participants' descriptions of familiar and unfamiliar data and our observations of how participants engaged with these data. We now synthesise our findings, identifying different \textit{patterns of activities} and their related \textit{data attributes} involved in data-centric sensemaking. The sensemaking efforts which we observed can be synthesized into three clusters of activities: \textit{inspecting} the data, \textit{engaging} with the data content more deeply and \textit{placing} data within broader contexts (Figure \ref{fig:activitycluster}). We also examine the relation of the clusters of sensemaking activities to information structures needed for reuse and discuss three emergent themes in the context of this synthesis. Here, we define: 
\begin{itemize}
\item \textit{Activity patterns} as the actions, both physical and cognitive, which people undertake when making sense of data
\item \textit{Data attributes} as characteristics of the data which people interact with as they perform a set of activities
\item \textit{Clusters} as the activity patterns, with their related attributes, which tend to occur together
\end{itemize}

\textbf{C1, inspecting,} contains activities and attributes that provide participants with a broad overview of the data, such as understanding the data's general topic, title, structure and format. In the \textit{unknown} data, we observed that most participants scanned the spreadsheet first vertically to look at the number of rows and to get an idea of missing values and then horizontally to look at the headers. \newline

\textbf{C2} represents a deeper \textbf{engagement with the content of the data}, including activities such as establishing relationships between columns, performing simple analyses, picking out examples of particular values, conducting quality assessments and trying to understand uncertainties attached to the data, by questioning, e.g. the meaning of missing values or abbreviations and acronyms. \newline

In \textbf{C3}, we observed participants \textbf{placing} data in relation to the world and different contexts. They worked to understand how the data were related to study designs, to disciplinary norms as well as to temporal and geographic considerations to understand the representativeness of the content. They questioned, e.g., the level of detail (granularity) presented in the data as well as the data's original purpose.\newline

Our findings show that level one (C1) of Figure \ref{fig:activitycluster} was mostly done alone; level two and three (C2, C3) were often solved in collaboration. When participants described their own as well as our data, critical details emerged only after the initial description, when both the interviewer and the participant could see and interact with the data together. These conversations moved away from objective descriptions towards describing the complexity of qualitative judgements behind the (quantitative) variables, as well as to rich descriptions of factors influencing the origination of data.
This echoes literature in critical data studies, conceptualising data as the product of socio-technical arrangements but also as a medium through which conversation and negotiation can occur \citep{DBLP:journals/bigdata/NeffTFO17}.

When discussing their data, participants made use of the information structures identified in Table \ref{tab:supplemental_things} and their related qualities (Table \ref{table:distancetothedata} and \ref{table:methodologicalapproach}) across all dimensions as they worked to make their data understood. Many also referenced the lack of contextual information (e.g. purpose, collection methods) in the \textit{unknown} data as being a stumbling block to understanding. 

The importance of needing contextual information to support data reuse, at both the level of the data \citep{borgman2015big,DBLP:conf/jcdl/FanielKKBY13,DBLP:conf/asist/FanielKY12} and of digital collections \citep{DBLP:journals/ijdc/BakerY09,Chin:2004:CSC:1031607.1031677,DBLP:journals/jd/Lee11}, has been extensively noted in the literature. Our work is in line with these findings, particularly in noting the importance of information describing data collection conditions and methodological details. Recent work (\citet{faniel2019context}) also draws attention to descriptions of what we term ``encodings" in Table \ref{table:quotetable}, to describe the codes that participants create and use when working with data. 

\citet{koestendataset} provide a summary of the literature examining particular data attributes for reuse, listing which papers document the importance of certain data and documentation characteristics. While our findings add to this literature, particularly by presenting important attributes along the lines of an individual’s knowledge of the data and its creation process (Table \ref{table:distancetothedata}) and the methodological narrative (Table \ref{table:methodologicalapproach}), our primary aim is not to isolate data attributes needed for reuse, as much work has already investigated this problem. Rather, the work we present here takes the lens of analysing how those attributes are brought together by the activities they afford, which we group to patterns of sensemaking activities. 

Translating these findings into interaction guidance and subsequently into tools supporting reuse presents a challenge, in part because of the dynamics and context-specific nature of working with data \citep{DBLP:conf/chi/KrossG19,DBLP:conf/chi/MullerLWPTLDE19}. The in-depth descriptions of study set-ups, purposes of data collection and domain specific knowledge brought by our participants underscores this problem. As a way to address this challenge, Figure \ref{fig:activitycluster} 
can be viewed in the context of work using design patterns in areas such as software engineering~\citep{gamma1995design}, user interface design~\citep{granlund2001pattern}, or ontology design~\citep{gangemi2009ontology}. 

This approach identifies high-level patterns as a way to provide repeatable solutions to recurring design problems. This creates possibilities for a level of formalisation that enables the development of flexible designs and tools. Our results are in line with \citep{Boukhelifa:2017:DWC:3025453.3025738,DBLP:conf/chi/KoestenKTS19,marchionini2005accessing}, who see flexibility as being key to supporting real-world data workflows. Figure \ref{fig:activitycluster} therefore represents a patterns-based approach to conceptualising the processes involved in the initial stages of data-centric sensemaking. \newline

To further contextualise Figure \ref{fig:activitycluster} and to illustrate how our findings could spur design efforts, we discuss three specific themes that emerged in our research at the level of each identified cluster. For each theme, we present design recommendations. The recommendations we propose exist in parallel to research in information visualization \citep{DBLP:conf/vl/Shneiderman96}, which suggests visual support for a high-level taxonomy of data tasks and types. Our study brings a deeper perspective to understanding information needs focusing on structured data, suggesting a wider variety of data-related tasks undertaken by users, which may or may not be supported through visual exploration. Our recommendations build on what we have learned about researchers’ sensemaking activities and workflows; our aim is to disrupt these workflows as little as possible. We therefore propose functionalities and approaches to support sensemaking that could be integrated within analysis tools already used by researchers, such as common programming languages or libraries. 

\subsection{Understanding shape}
When inspecting a dataset for the first time, see Cluster 1, participants either discussed the data in a linear fashion, addressing each attribute individually before moving to the next, or they took a more interwoven approach, mixing descriptions of dataset attributes with analyses and questions. This interwoven approach also has overlaps with activities in the second cluster of Figure \ref{fig:activitycluster}. 

As they engaged in inspecting acitvities, participants aimed to arrive at an overview, to create a high-level representation of the entire dataset in their head while engaging with it (see also \citep{DBLP:conf/chi/KoestenKTS17}). We observed different levels of focus in this process. Participants alternated between ``zooming out" to describe the data at the level of the entire spreadsheet, e.g. the number of observations or format of the data, and ``zooming in" to look at specific cell values or individual parts of the data. Participants adopting a more interwoven approach tended to engage in the process of zooming in and out more often than those using a linear approach.

This desire to understand the data as a whole has parallels in the information science literature, where the need to understand an entire information collection at a high level has been mentioned \citep{rieh2016towards}. Discussing the visual aspects of sensemaking, \citet{DBLP:conf/hvei/Russell03} also mentions the need to understand what is in a whole collection. \citet{white2009exploratory} recommends allowing users to filter, sort and explore different views of information on demand for complex search tasks. In our study, the information is distributed among the cells of the dataset, the structure and organization of the data, as well as any related information structures.

\subsubsection{Recommendations}
\label{subsubsec:recommendationsshape}
Understanding the shape of a dataset can be supported through interface design and functionalities in a number of ways. Our results show that data needs to be understood as a whole, on the \textbf{level of the entire dataset}. This suggests summarization methods, which can be of textual, visual or statistical nature, that provide a zoomed-out view of the data (e.g. \citet{DBLP:journals/corr/abs-1805-03677}). At the same time, participants also engaged with subsets of the data, particularly individual columns; these patterns could be supported through zooming in via \textbf{column level summaries}, including interactive plots and visualizations at the column level (e.g., this idea is partially realised by Kaggle in their dataset previews\footnote{\url{https://www.kaggle.com/}}). Future research could look at different ways of expressing a column-based notion of provenance, such as where the data in a column comes from, how it was created or from where it was derived. Given the importance of scanning and zooming in and out (as mentioned in literature such as \cite{DBLP:conf/vl/Shneiderman96}), data search engines and displays should optimize this functionality to make these processes as fast as possible; including horizontal scrolling to accommodate spreadsheets with more columns.

Similarly, certain types of information structures attached to the dataset facilitate particular sensemaking patterns over others. A README file with a summary of the dataset's size and format may provide the information necessary for a zoomed out inspection of the data; an interactive map of the area where a specimen was collected may be more suitable to a zoomed in approach, as well as enabling the activities described in Cluster 2.

\subsection{``Strange things" as an entrypoint, not an obstacle} 
Participants repeatedly encountered and dealt with ``strange things" in both data sources, i.e. outliers, errors, missing data, and inconsistencies in formatting. As they wrestled with the unexpected in the data, they engaged in the patterns identified in Cluster 2, such as expressing uncertainty, seeking relationships or performing analyses. 

Whereas \citep{doi:10.1002/asi.24221} describe dealing with conflicts as a barrier to sensemaking, our findings suggest that conflict is a useful and accelerating moment in the exploration of data. The concept that real data is usually messy and complex was internalised by our participants. Participants were neither surprised nor alienated by conflicting data; in contrast, errors and uncertainties were expected and participants applied different analytical strategies to overcome them, a finding also in line with recent literature \citep{Boukhelifa:2017:DWC:3025453.3025738,DBLP:journals/bigdata/NeffTFO17, DBLP:conf/chi/KoestenKTS17}.
 Participants repeatedly emphasized the need to communicate information about sources of error and possible uncertainties to potential data consumers, although there were a variety of communication methods used to do so, some of which are detailed in Table \ref{tab:supplemental_things}. Methods for communicating information about strange things in the data were sometimes chosen arbitrarily or convenience-based. Some of this information was embedded within data themselves, leading to potential problems in machine readability. Others were not linked to the data in a sustainable way, making them unsuitable for long-term preservation of meaning. 

\subsubsection{Recommendations} 
Our findings suggest that errors can be seen as an entry point to sensemaking, as flags to investigate further. This provides an interesting direction to explore for sensemaking functionalities in tools. Rather than flattening out data by making it cleaner, tools could instead \textbf{flag and highlight strange things} to make users more aware of their presence. Column summaries, as mentioned in \ref{subsubsec:recommendationsshape}, could include explanations of abbreviations and missing values, metrics or links to other information structures necessary for understanding the column's content. Datasets should include \textbf{links to basic concepts} (used in the data or in the documentation) such as common practices in code documentation or ``the web" (i.e. in Wikipedia / Wikidata) to provide context. Documentation about the narrative surrounding these strange things should also be more standardised and linked directly to these flags in a sustainable way. 

Other sensemaking patterns identified in C2 can be supported by customised interactive visualisations. Displaying the entire data, as described in \ref{subsubsec:recommendationsshape}, but \textbf{highlighting relationships between columns or entities} could allow users to more easily pick up relationships between columns. Tools could also \textbf{display trends and patterns} extracted from the dataset and allow users to select those attributes of the data that are of interest. Following this idea, data producers could identify anchor variables, those which they consider most important in their dataset; this could further aid sensemaking activities by focusing summarisation efforts.

\subsection{Perspectives in placing}
Participants place data and their representativeness in a range of broader contexts (the world, disciplinary norms, methodological contexts of creation). While we present these placing activities separately in Cluster 3, they can in fact be closely related. We saw this particularly in how participants placed data in terms of a study's methods and their own disciplinary expertise. 

Details about data creation are often implicit within a domain's epistemic norms \citep{leonelli2016data}. Even with the best documentation, this complicates cross-disciplinary data reuse. A data consumer from another domain may not have the experience necessary to understand or evaluate the appropriateness of a particular methodological approach. Additionally, our participants' concept of the details needed for reuse encompassed much more than just a step-by-step process of how a study was conducted. Rather, for both quantitative and qualitative data (see Table \ref{table:methodologicalapproach}), participants needed details about the entire narrative surrounding data creation, i.e. why a certain method was chosen or the unique, local aspects about a study's set-up and their attached constraints. 
This need for expanded and robust methodological narratives mirrors recent calls for details beyond those provided by standardized metadata and reporting conventions, particularly for the reuse of qualitative data \citep{doi:10.1177/1609406919868870}.
\newline
\indent We also found that the granularity of these narratives is related to a potential data consumer's expertise or distance to the data, with experts needing more detailed information about study descriptions. Table \ref{table:distancetothedata} also shows common attributes, aside from methodology, that are important in facilitating understanding, independent of a data consumer's expertise with data, i.e. needing information about study objectives, usage restrictions, and explanations of categories and acronyms. 

\subsubsection{Recommendations}
Our findings highlight the need for flexible designs to support placing activities across the three identified levels of placing: the world, disciplinary norms and the study-set-up. Rather than designing for a specific type of user, tools should be designed to embrace different levels of expertise, allowing a potential data consumer to \textbf{drill down to the desired level of detail}. Semantic technologies \citep{Balog2018} also could be used to \textbf{link to standardized definitions} of disciplinary acronyms or terms, mirroring our recommendation in \ref{subsubsec:recommendationsshape} to link to external knowledge bases. Geographic information could be linked to a map or country registry to allow judgements of representativeness; a similar approach could be taken for certain disciplinary standards and study set-ups, such as standard experiment conditions, expected result ranges or commonly used confidence levels. Data citations, in particular their associated metadata, can contain detailed provenance information needed for sensemaking, offering another emerging possibility for providing the necessary context for data reuse \citep{groth2020fair}.\\

Our findings across all dimensions emphasize the collaborative nature of data-centric sensemaking and the omnipresent role of information structures throughout the sensemaking and reuse process. It has been suggested that the production and consumption of academic writing can be conceptualized as a form of dialogue (e.g. \citet{doi:10.1177/1474022211398106}); the broader practice of reusing data could itself be seen as a form of collaboration or conversation between data producers and consumers. The data producer must communicate the many (often collaborative) decisions which influence the creation of a dataset \citep{DBLP:journals/tvcg/MahyarT14,DBLP:journals/bigdata/NeffTFO17} to potential data reusers. 

A combination of focused documentation practices integrating different media types, together with prescribed interaction flows tailored to the sensemaking practices of both data producers and consumers, could facilitate the conversation implicit in reusing data. These could include solutions with adaptable data representations suited to varying levels of expertise and needs. 

\section{Study Limitations}
Although they were working in a wide range of disciplinary domains and research related roles, our sample population consisted of a particular type of professional: researchers who have published an article indexed in the Scopus database\footnote{\url{https://www.scopus.com}}. 

Scopus comes with a skew towards certain research fields; the Arts \& Humanities, for instance, are not as well-represented. Scopus has an extensive review process for the journals which it selects for inclusion \citep{elsevier2020scopus}; and there are roughly an equivalent number of journals from the broadly defined fields of social sciences, health sciences and physical sciences \citep{elsevier2020scopus}. While the limitations of Scopus could lead to a potential bias in our sample, the selection criteria we applied also ensured that the sample population met our study requirement of speaking with different types of researchers with data experience.

As the study was conducted with researchers, our findings may not be directly applicable to other individuals. Focusing on researchers met the goals of our study, particularly our aim of examining sensemaking in light of reuse. However, we believe the general sensemaking patterns emerging from this study are to some extent transferable between different skill sets; simply the execution of how these goals are achieved might look different for people with a higher or lower level of data literacy. Nonetheless, the study would need to be repeated with different populations in order to apply our findings more broadly. 

Our participants work in a variety of countries; English was not the native language for all. To account for this, we selected our sample from those responding to our recruitment messages carefully to ensure that participants had a high degree of English fluency. While we see the global spread and disciplinary diversity of our sample as a strength of the study, we also recognise that data, research, and sensemaking practices are influenced by social, legal, and economic contexts unique to both country and disciplinary domains. 

The sensemaking patterns which we identify could also be limited by the data themselves. Different data may have surfaced different data attributes. By including participants' data, as well as the \textit{unknown} dataset in the study, we attempted to balance this potential bias. Another potential limitation is that describing their data first might have primed participants for performing the verbal summarisation task with the \textit{unknown} data, influencing the way that they performed the second task. Given that any data description will be based on a particpants' prior experience, we believe this is a natural side effect of these types of studies.

Finally, it is important to note that we intentionally did not ask participants to bring metadata for their \textit{known data} as we wanted to see what types of data, metadata and other contextual information they felt that they needed to bring without being prompted. Similarly, the study design allowed us to capture what participants felt was missing from the \textit{unknown data} and to identify what additional information was needed. This is especially relevant as data does not always come with complete or accurate or meaningful metadata or enough information for reuse, as detailed in the background section.

\section{Conclusions}

In this study, we investigated common patterns in sensemaking activities in initial encounters with data, particularly in light of potential data reuse.

We identified three clusters of activities involved in initial data-centric sensemaking (\textit{inspecting, engaging with content, and placing in context}), and detailed the observed activities and data attributes relevant in these clusters. This approach provides an avenue to bring focus to design efforts, narrowing down the  many technologically feasible solutions to those specifically supporting the sensemaking needs of data consumers. To summarize, the contributions of the paper are: 
\begin{itemize}[noitemsep]
\item activity patterns for data-centric sensemaking;
\item a framework of these patterns and their associated data attributes;
\item user needs for data reuse;
\item design recommendations to support the identified activity patterns.
\end{itemize}

Our work illustrates the large space for future research trajectories in this area to validate and apply insights within different contexts of data-centric work practices. This could include investigating the identified activity patterns with different data or with individuals working outside of research. Other work could focus on how to apply the detailed insights and recommendations to existing user workflows. Such work could focus on determining the best way to present and allow interaction with data to facilitate sensemaking. Similarly, such work could explore the integration of our findings into existing services and platforms, particularly with regard to multidisciplinary data. 

Sensemaking allows individuals to create rational accounts of the world which enable action \citep{maitlis2005social}. In this work, data-centric sensemaking enables a particular type of action: the reuse of data in research. Understanding how people make sense of data, and exploring designs to support these practices, therefore, plays a key role in realizing the potential of data reuse. 

\label{Conclusion}

\section*{Acknowledgements}
This research is partially supported by the Data Stories project, funded by EPSRC research grant No. EP/P025676/1 and by the NWO Grant 652.001.002 Re-search: Contextual search for scientific research data.

\label{}



\bibliographystyle{elsarticle-harv} 
\bibliography{sample-bibliography}





\end{document}